\documentclass[reprint,superscriptaddress,amsmath,amssymb,aps,prc]{revtex4-2}

\usepackage{soul} 
\usepackage{graphicx}
\usepackage{xcolor}
\usepackage{booktabs}
\usepackage{dcolumn}
\usepackage{bm}
\usepackage{graphics}
\usepackage{svg}
\usepackage{mathtools}
\usepackage{siunitx}
\usepackage{longtable}

\newcommand{\todo}[1]{\textcolor{red}{\textbf{TODO:} #1}}

\begin{document}
\newcommand {\urss}[1]{\ensuremath{_{\mathrm{#1}}}}
\newcommand {\td}[1]{\protect \todo{#1}} 
\newcommand {\nistb}{\affiliation{National Institute of Standards and Technology, Boulder, CO, 80303}}
\newcommand {\nistg}{\affiliation{National Institute of Standards and Technology, Gaithersburg, MD, 20899}}
\newcommand {\nistag}{\affiliation{Associate, National Institute of Standards and Technology, Gaithersburg, MD, 20899}}
\newcommand {\cu}{\affiliation{The University of Colorado, Boulder, CO, 80309}}
\newcommand {\clemson}{\affiliation{Department of Physics and Astronomy, Clemson University, Clemson, SC 29634}}
\newcommand {\lum}
{\affiliation{Loyola University Maryland, Baltimore, MD 21210}}
\newcommand {\mi}{\affiliation{Massachusetts Institute of Technology, Cambridge, Massachusetts 02139}}
\newcommand {\cfa}
{\affiliation{Center for Astrophysics | Harvard and Smithsonian, Cambridge, MA 02138}}
\newcommand {\nasa}{\affiliation{CRESST-II and High Energy Astrophysics Branch, NASA/GSFC, Greenbelt, MD 20771}}
\newcommand {\umd}{\affiliation{Department of Astronomy, University of Maryland, College Park, MD 20742}}
\newcommand {\auburn}{\affiliation{Department of Physics, Auburn University, Auburn, AL 36849}}

\preprint{APS/123-QED}

\title{Benchmark Ne-like Transitions of W, Re, Os, Ir, and Bi Measured by \\ High-resolution TES x-ray Spectroscopy with Absolute Energy Calibration}


\author{G. Mondeel}
\clemson
\mi
\email[Contacts: ]{ gmondeel@mit.edu, etakacs@clemson.edu}

\author{C.A. Faiyaz}
\clemson

\author{Dipti}
\clemson
\nistag

\author{H. Staiger}
\clemson
\nistag

\author{L. Brent}
\lum
\nistag

\author{A.C. Gall}
\auburn
\cfa

\author{C.I. Szabo}
\nistg

\author{G. O'Neil}
\nistb

\author{D.S. Swetz}
\nistb

\author{P. Szypryt}
\nistb
\cu

\author{J.W. Dean}
\nistb
\cu

\author{Yu. Ralchenko}
\nasa
\umd

\author{E. Takacs}
\clemson
\nistag

\date{\today}

\begin{abstract}

We present high-resolution measurements of $n=2–3$ transitions in Ne-like tungsten, rhenium, osmium, iridium, and bismuth using a transition-edge sensor (TES) microcalorimeter. These measurements provide new benchmarks for Ne-like ions in the high-Z regime ($Z > 74$), extending previous EBIT studies of the sequence. To achieve sub-eV accuracy, we developed a dedicated x-ray fluorescence source employing well-characterized K-shell emission lines from metallic elements.
Our results offer stringent tests for relativistic atomic structure theories and demonstrate the advantage of TES spectroscopy for precision measurements in highly charged ions.
\end{abstract}

\maketitle

\section{\label{intro}Introduction}

High-precision measurements of x-ray transitions in highly charged ions provide critical benchmarks for atomic structure theory, particularly in regimes where relativistic and quantum electrodynamics (QED) effects are significant \cite{Indelicato2019}. Among these, the Ne-like isoelectronic sequence characterized by a closed-shell $1s^2 2s^2 2p^6$ ground configuration serves as a fertile testing ground for theory due to its relatively simple structure and the sensitivity of its $n=2 - 3$ transitions to electron correlation and QED corrections.

Early observations of $n=2 - 3$ transitions in Ne-like ions were conducted using a variety of high-temperature plasma sources, including laser-produced plasmas \cite{Burkhalter1975, Gordon1979, Gordon1980, Conturie1981}, tokamaks \cite{Klapisch1978, Kallne1983}, z-pinch \cite{Burkhalter1979}, exploding wire \cite{Burkhalter1977}, vacuum spark \cite{Aglitsky1981}, and solar atmospheric emissions \cite{Parkinson1973FeXVII, Hutcheon1976FeXVII, McKenzie1980SolarFlare, Phillips1982SolarMax}. These pioneering studies demonstrated the presence of strong line radiation from Ne-like ions across a wide range of atomic numbers and plasma conditions, establishing their diagnostic utility. Notably, x-ray spectra of Ne-like ions up to xenon ($Z=54$) were recorded in laser-produced plasmas \cite{Conturie1981} and beam-foil experiments \cite{Dietrich1985XeQuadrupole}, providing early benchmarks for atomic structure theory. The latter method was the first to report electric quadrupole (E2) transitions in these systems \cite{Dietrich1985XeQuadrupole}.

These efforts laid the groundwork for more systematic and high-resolution measurements, pioneered by Beiersdorfer \textit{et al.} using a tokamak device~\cite{Beiersdorfer1986}, which enabled the first observations of a magnetic quadrupole (M2) transition and the identification of satellite lines from adjacent charge states in Ne-like silver ($Z=47$). These measurements were later extended, providing systematic measurements of $n=2 - 3$ transitions in Ne-like ions from Ag$^{37+}$ through Xe$^{44+}$ and rare-earth elements such as La$^{47+}$, Nd$^{50+}$, and Eu$^{53+}$~\cite{Beiersdorfer1988}.
The polarization of the M2 transition was subsequently measured in Ne-like Ba$^{56+}$~\cite{Takacs1996}, offering insight into magnetic sublevel-specific excitation and revealing the influence of cascade feeding and resonant processes on line polarization.

Subsequent studies using low-inductance vacuum sparks~\cite{Aglitskii1986} and electron beam ion trap (EBIT) spectroscopy~\cite{Levine1989, Beiersdorfer1990_Yb, Beiersdorfer1990_EBIT, Beiersdorfer1993, Nakamura1999, Nakamura1999_XRayLasers, Nakamura2000_PRA} significantly expanded the dataset of Ne-like transitions across a broader range of atomic numbers ($Z=36 - 92$). These investigations provided valuable benchmarks for theoretical models, revealing smooth trends in transition energies and persistent discrepancies with multiconfiguration Dirac-Fock (MCDF) calculations—particularly for transitions involving 2s-hole states~\cite{Beiersdorfer1988, Beiersdorfer2012}. Such discrepancies underscore the importance of accurate theoretical treatments of electron correlation and QED effects in multielectron systems.

More recently, Brown \textit{et al.}~\cite{Brown2008} employed an x-ray microcalorimeter to study 2p-3d transitions in multiple charge states of gold, including the observation of Ne-like spectral lines. In parallel, Beiersdorfer \textit{et al.}~\cite{Beiersdorfer2012} explored the diagnostic potential of Ne-like tungsten in fusion plasmas \cite{Safronova2009}.

In an important development, Ne-like transitions were investigated near the avoided crossing of the $(2p^{-1}_{1/2} \, 3d_{3/2})_{J=1}$ and $(2s^{-1}_{1/2} \, 3p_{1/2})_{J=1}$ levels in the vicinity of Z=68~\cite{Beiersdorfer2016}. This work built upon earlier studies by Nakamura \textit{et al.}~\cite{Nakamura2000_PRA}, who observed a similar avoided crossing between the $(2p^{-1}_{1/2} \, 3s_{1/2})_{J=1}$ and $(2p^{-1}_{3/2} \, 3d_{5/2})_{J=1}$ levels near $Z = 55$. Both studies revealed strong configuration mixing between levels of identical total angular momentum and parity, leading to dramatic changes in oscillator strengths and transition energies across the crossing. These findings highlight the importance of accurately modeling configuration interaction and relativistic effects in highly charged ions~\cite{Safronova2001}. 

A follow-up study focused on europium (Z=63)~\cite{Beiersdorfer2019}, an element just below the crossing region. However, experimental data for Ne-like ions in the high-$Z$ regime ($Z > 74$) remain sparse. This region is particularly compelling because the avoided crossings have stabilized, making it an ideal domain for testing the convergence of theoretical models. As shown in Figure 1, the evolution of transition probabilities and scaled upper level energies across the Ne-like sequence reflects the cumulative impact of configuration mixing, relativistic effects, and QED corrections \cite{Safronova2001}. Figure 1 spotlights how these effects manifest as abrupt changes in line strengths and level ordering near the crossing regions, complicating line identifications and theoretical modeling. 

The high-$Z$ region beyond these avoided crossings provides benchmarks for atomic structure calculations where relativistic, Breit, and QED contributions become increasingly important~\cite{Beiersdorfer2016}. In this work, we present the first high-resolution measurements of $n=2$–$3$ transitions in Ne-like rhenium (Re$^{65+}$), osmium (Os$^{66+}$), and iridium (Ir$^{67+}$) using a transition-edge sensor (TES) microcalorimeter. We also report Ne-like W$^{64+}$ and Bi$^{73+}$ transition energies that extend the earlier measurements in~\cite{Beiersdorfer2012} and~\cite{Dietrich90}, respectively. This technique offers sub-eV energy precision in the 8–12 keV x-ray range and enables the resolution of individual lines and satellite features that were previously blended or inaccessible. Our measurements extend the EBIT-based Ne-like sequence into an unexplored $Z$ regime and provide critical benchmarks for theory in the post-avoided-crossing region.

\begin{figure}
    \centering
    \includegraphics[width=0.99\linewidth]{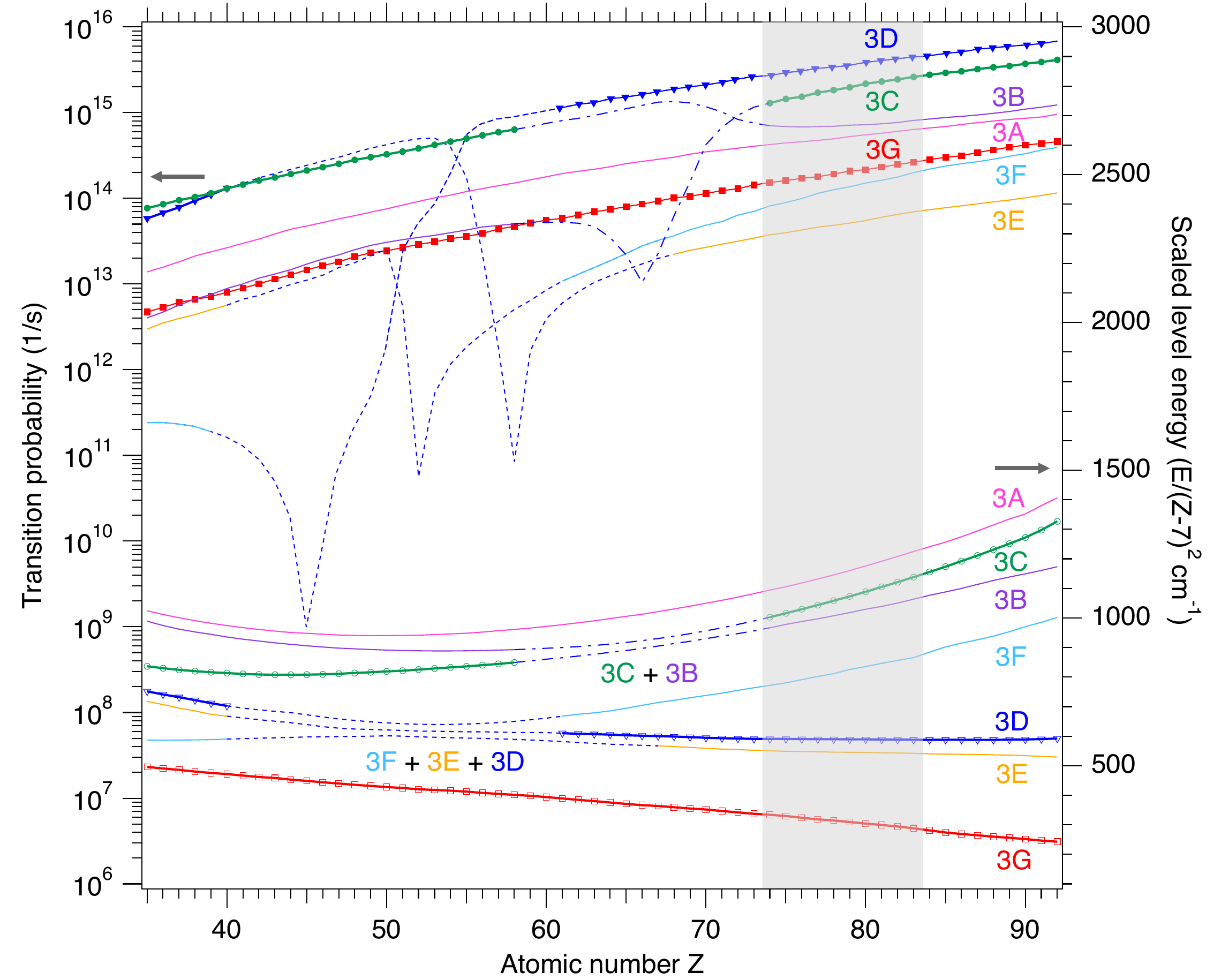}
    \caption{Transition probabilities (left axis) and scaled upper-level energies (right axis) of the main Ne-like E1 highly charged ion transitions along the isoelectronic sequence \cite{Safronova2001}. The figure illustrates strong configuration mixing among the odd-parity $J=1$ upper levels across the upper half of the periodic table. As the energy levels pass through avoided crossings along the isoelectronic sequence, they exchange their dominant configuration character. The transition probabilities also exhibit dramatic variations.}
    \label{Crossing}
\end{figure}

To ensure accurate energy calibration, we developed a dedicated x-ray fluorescence calibration source based on the National Institute of Standards and Technology (NIST) Standard Reference Database (SRD) 128 \cite{SRD128}, a direct outcome of the comprehensive evaluation by Deslattes \textit{et al.} \cite{Deslattes2003}. This source includes K emission lines from cobalt, copper, zinc, and germanium, which were originally measured using vacuum double-crystal spectrometers with traceability to the SI meter via optical interferometry \cite{Szabo2022}. By anchoring our calibration to these well-characterized lines, we achieve absolute energy uncertainties at the level of 0.8 eV, ensuring that our measurements are directly comparable to theoretical predictions.

To interpret our results, we employ the FAC~\cite{Gu2008} and GRASP2K ~\cite{Jonsson2013} atomic structure codes. The former is a widely used relativistic-model-potential code for calculation of structure and collisional data in simulations of spectra from highly charged ions. The latter implements the multiconfiguration Dirac-Hartree-Fock (MCDHF) method in a fully relativistic framework~\cite{Jonsson2013}. GRASP2K allows for the inclusion of electron correlation, Breit interaction, and leading QED effects, making it well-suited for modeling the complex structure of high-$Z$ Ne-like ions. By comparing our measured energies with GRASP2K predictions and with earlier EBIT data, we assess the current state of theoretical accuracy and identify areas where further refinement is needed.

\section{\label{experiment}Measurement Details}

\subsection{\label{ebit}NIST EBIT}

The NIST EBIT is designed for the production and confinement of highly charged ions (HCIs) through successive electron impact ionization. It operates by compressing a high-current electron beam (up to 150~mA) using a strong magnetic field (approximately 2.7-3~T), focusing the beam to a narrow diameter of about 60~$\mu$m. Axial confinement is achieved via a set of cylindrical drift tubes, with voltages configured to trap ions along the beam axis. The actual electron beam energy, nominally up to 30~keV, is slightly reduced due to space-charge effects and partial neutralization by trapped ions. Ion injection is typically performed using a Metal Vapor Vacuum Arc (MeVVA) source, allowing for controlled introduction of singly charged ions into the trap.

The NIST EBIT has played a fundamental role in advancing atomic physics of highly charged ions, enabling precision spectroscopic measurements, tests of quantum electrodynamics, and benchmarking atomic structure theory~\cite{Morgan95, Chantler1999, Chantler2000,Payne2014,Chantler2012,Chantler2014}. It has been used to measure fine-structure transitions, investigate magnetic-dipole (M1) and quadrupole (M2) transitions in high-$Z$ elements with plasma diagnostic potential~\cite{Takacs2006, Ralchenko2006, Ralchenko2008, Ralchenko2011} and study electron impact and anisotropic excitation processes~\cite{Ralchenk2013, Yang2025, Yang2025b}. 

For the spectra shown in Fig.~\ref{fig:exp_theo}, the nominal electron beam energies were 15.6~keV for W, 19.0~keV for Re, 20.0~keV for Os and Bi, and 19.1~keV for Ir. The corresponding beam currents were approximately 75~mA, 110~mA, 90~mA, and 110~mA, respectively.

\subsection{NIST EBIT TES Array x-ray Detector}

\begin{figure}
    \centering
    \includegraphics[width=0.99\linewidth]{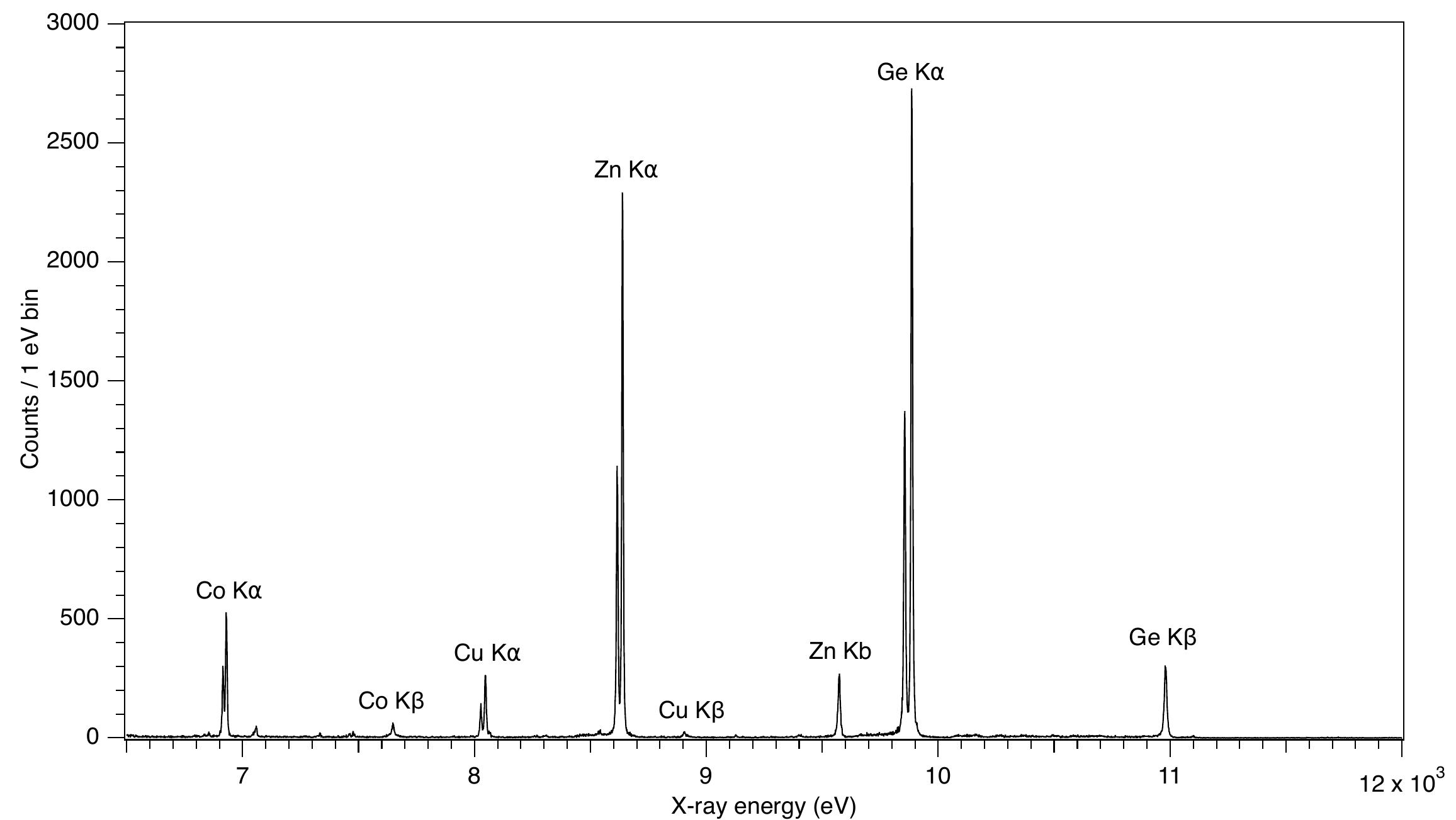}
    \caption{Measured fluorescence spectrum used to calibrate the TES instrument. Reference line energies were taken from NIST SRD 128 \cite{Deslattes2003, SRD128}.}
    \label{fig:calibration_lines}
\end{figure}

The NIST EBIT TES array x-ray detector comprises 192 cryogenic TES detectors each acting as an independent spectrometer pixel designed for high-resolution x-ray spectroscopy~\cite{szypryt_transitionedge_2019} in the energy range concerned. The cryostat that maintains the regulating temperature of approximately 60 mK is an adiabatic demagnetization refrigerator (ADR) with a hold-time of roughly 20 hours. Each pixel incorporates a superconducting MoCu bilayer film (\SI{65}{\nano\meter} Mo / \SI{215}{\nano\meter} Cu) with a critical temperature of 110~mK, coupled to a \SI{1}{\micro\meter} thick gold absorber optimized for efficient x-ray energy capture. The detectors are voltage-biased to place them in the superconducting transition.

Photons are only incident on the gold absorber due to an aperture array placed above the detectors. Upon absorption of an x-ray photon, the resulting temperature rise drives the TES further into its resistive transition region. This shift produces a transient decrease in current ---referred to as a \emph{pulse}--- whose amplitude is monotonic with the photon energy. While the NIST EBIT TES array has traditionally been used in the 1 keV - 7 keV x-ray energy region \cite{ONeil2020_PRA, Yang2023, Hosier2023, Burke2024, Faiyaz2025}, this work demonstrates its capability in the high-energy regime, with reliable performance at photon energies exceeding 10~keV. Despite the reduced absorber efficiency at higher energies, the detector maintains sub-eV precision and achieves a full-width-at-half-maximum (FWHM) energy resolution of 5~eV. This enables the resolution of fine spectral features in highly charged ions, including satellite lines and transitions that were previously inaccessible due to blending or instrumental limitations.


\begin{figure}[t]
\centering
 \includegraphics[width=0.99\linewidth]{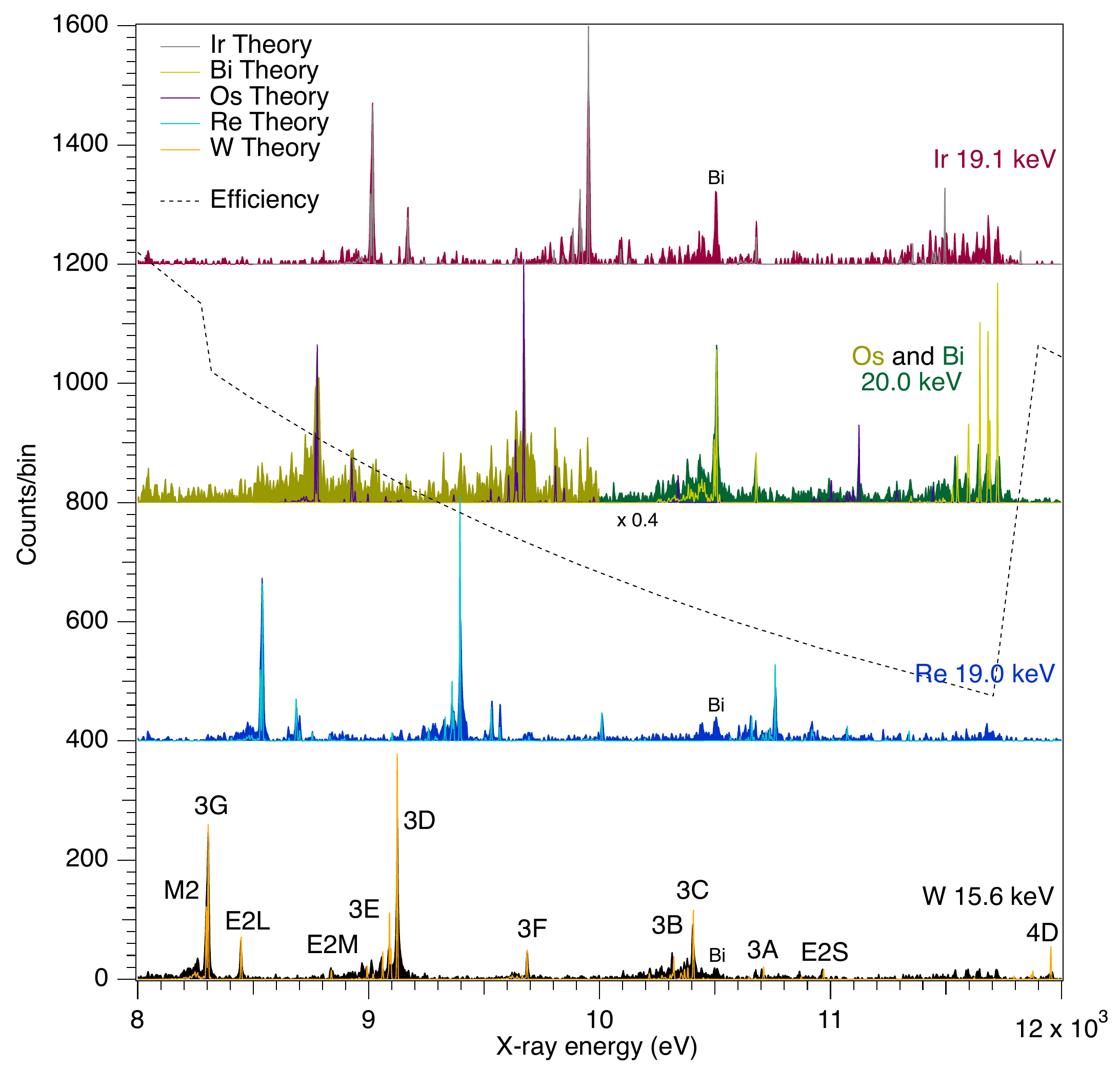}
\caption{Theoretical and experimental TES x-ray spectra of W, Re, Os, Ir, and Bi with dominant spectral lines arising from Ne-like and neighboring charge states. Bi appeared as a contaminant from the MeVVA ion source in all spectra and was particularly prominent in the Os spectrum. The approximate relative detector efficiency, shown as a dashed line, was applied to the theoretical line intensities for comparison with the measured spectra. Each theoretical spectrum was normalized to the measured 3G line intensity.}
\label{fig:exp_theo}
\end{figure}

\subsection{\label{calibration}x-ray Fluorescence Source for Calibration}
To ensure accurate energy measurements of x-ray transitions in highly charged ions in the 10 keV region, the TES array at the NIST EBIT is calibrated using a dedicated external x-ray source. During the EBIT acquisitions, the fluorescence source was operated in cycles of 1 min on and 4 min off. Calibration in this energy region is particularly important for studies involving K- and L-shell transitions in mid- to high-$Z$ elements.

The calibration system includes a commercial x-ray tube that generates bremsstrahlung and characteristic radiation by accelerating electrons toward a palladium target. This primary radiation excites fluorescence in a secondary target composed of carefully selected metallic samples mounted on an aluminum ring. These samples emit well-known characteristic x-ray lines, which are detected by the TES and used to construct a calibration curve. The setup is housed under vacuum to maintain clean conditions and minimize background.

For each x-ray absorption event the TES spectrometer records a pulse as the energy of the photon raises the temperature of the TES before returning to the bath temperature. These pulses are not directly traceable to a definition of energy. Therefore, calibration of TES spectrometers is essential, and a limiting factor in their ability to make sub-eV accurate measurements. This issue is compounded by the fact that the detector response is slightly nonlinear with energy. To extend the NIST EBIT TES measurements to higher energies, we developed a dedicated fluorescence calibration sample. The elements included and the lineshape reference used are: Ge, Zn, and a custom alloy of Sc~\cite{Dean2019,Dean2020}, V~\cite{Chantler2006}, Mn~\cite{Holzer1997}, Co~\cite{Holzer1997}, and Cu~\cite{Mendenhall2017}. The x-ray lines from these elements are traceable to the SI definition of the meter through crystal spectrometer measurements and the 2003 redetermination of x-ray lines by Deslattes~\textit{et al.}~\cite{Deslattes2003}.

\begin{table*}[]
\centering
\caption{Measured Ne-like transition energies to the $2s^22p^6$ ground state in eV for W, Re, Os, Ir, and Bi.
Numbers in parentheses denote statistical and fitting uncertainties,
while numbers in square brackets denote calibration uncertainties, all expressed in units of the last quoted digit.
All uncertainties are quoted at the $1\sigma$ level. The Counts/eV column gives the fitted peak height above background. GRASP and RMBPT~\cite{Safronova2001,SafronovaNIFS61} calculations and previous experimental
measurements are included for comparison (W:~\cite{Beiersdorfer2012}, Bi:~\cite{Dietrich90}).}
\label{tab:ne_like_all}

\small
\setlength{\tabcolsep}{3.5pt}
\renewcommand{\arraystretch}{1.05}

\begin{tabular}{llcccrcrrr}
\toprule
Element & Line & Core & Valence & $J$
& Experiment & Counts/eV & GRASP & RMBPT & Previous Exp. \\
\midrule

\textbf{W} ($Z=74$)
& M2
& $2s^2 2p^5_{3/2}$ & $3s_{1/2}$ & 2
& 8298.81(39)[12] & 148 & 8298.09 & 8297.57 & 8299.22(40) \\

& 3G
& $2s^2 2p^5_{3/2}$ & $3s_{1/2}$ & 1
& 8307.04(22)[12] & 293 & 8306.53 & 8306.01 & 8307.51(40) \\

& E2L
& $2s^2 2p^5_{3/2}$ & $3p_{1/2}$ & 2
& 8448.17(30)[10] & 68 & 8448.48 & & 8450.14(60) \\

& E2M
& $2s^2 2p^5_{3/2}$ & $3p_{3/2}$ & 2
& 8837.48(92)[14] & 17 & 8838.61 & & \\

& 3E
& $2s^2 2p^5_{3/2}$ & $3d_{3/2}$ & 1
& 8993.1(10)[02] & 10 & 8994.94 & 8993.53 & 8996.31(50) \\

& 3D
& $2s^2 2p^5_{3/2}$ & $3d_{5/2}$ & 1
& 9126.22(8)[27] & 184 & 9125.17 & 9125.35 & 9126.25(50) \\

& 3F
& $2s^2 2p^5_{1/2}$ & $3s_{1/2}$ & 1
& 9687.61(17)[16] & 35 & 9687.69 & 9687.35 & 9689.29(50) \\

& 3B
& $2s 2p^6_{1/2}$ & $3p_{1/2}$ & 1
& 10315.8(11)[03] & 26 & 10315.97 & 10316.09 & 10317.23(50) \\

& 3C
& $2s^2 2p^5_{1/2}$ & $3d_{3/2}$ & 1
& 10405.68(46)[40] & 70 & 10407.49 & 10406.33 & 10408.69(40) \\

& 3A
& $2s 2p^6_{1/2}$ & $3p_{3/2}$ & 1
& 10703.6(16)[05] & 9 & 10704.62 & 10704.12 & 10706.85(90) \\

& E2S
& $2s 2p^6_{1/2}$ & $3d_{5/2}$ & 2
& 10965.4(17)[07] & 9 & 10966.19 & & 10967.8(12) \\

\midrule

\textbf{Re} ($Z=75$)
& M2
& $2s^2 2p^5_{3/2}$ & $3s_{1/2}$ & 2
& 8532.77(30)[07] & 169 & 8532.80 & 8532.21 & \\

& 3G
& $2s^2 2p^5_{3/2}$ & $3s_{1/2}$ & 1
& 8541.32(11)[07] & 413 & 8541.40 & 8540.81 & \\

& E2L
& $2s^2 2p^5_{3/2}$ & $3p_{1/2}$ & 2
& 8687.57(30)[05] & 82 & 8686.37 & & \\

& E2M
& $2s^2 2p^5_{3/2}$ & $3p_{3/2}$ & 2
& & & 9101.67 & & \\

& 3E
& $2s^2 2p^5_{3/2}$ & $3d_{3/2}$ & 1
& & & 9260.69 & & \\

& 3D
& $2s^2 2p^5_{3/2}$ & $3d_{5/2}$ & 1
& 9399.07(41)[16] & 294 & 9397.01 & 9397.13 & \\

& 3F
& $2s^2 2p^5_{1/2}$ & $3s_{1/2}$ & 1
& 10011.59(48)[10] & 36 & 10011.17 & 10010.80 & \\

& 3B
& $2s 2p^6_{1/2}$ & $3p_{1/2}$ & 1
& 10653.81(85)[33] & 35 & 10653.04 & 10653.03 & \\

& 3C
& $2s^2 2p^5_{1/2}$ & $3d_{3/2}$ & 1
& 10763.01(51)[32] & 92 & 10761.81 & 10760.66 & \\

& 3A
& $2s 2p^6_{1/2}$ & $3p_{3/2}$ & 1
& 11068.85(77)[39] & 13 & 11066.49 & 11065.93 & \\

& E2S
& $2s 2p^6_{1/2}$ & $3d_{5/2}$ & 2
& & & 11336.50 & & \\

\midrule

\textbf{Os} ($Z=76$)
& M2
& $2s^2 2p^5_{3/2}$ & $3s_{1/2}$ & 2
& 8770.42(69)[09] & 33 & 8770.08 & 8769.41 & \\

& 3G
& $2s^2 2p^5_{3/2}$ & $3s_{1/2}$ & 1
& 8779.69(48)[10] & 47 & 8778.85 & 8778.18 & \\

& E2L
& $2s^2 2p^5_{3/2}$ & $3p_{1/2}$ & 2
& 8926.6(18)[2] & 10 & 8926.87 & & \\

& E2M
& $2s^2 2p^5_{3/2}$ & $3p_{3/2}$ & 2
& & & 9368.67 & & \\

& 3E
& $2s^2 2p^5_{3/2}$ & $3d_{3/2}$ & 1
& & & 9530.40 & & \\

& 3D
& $2s^2 2p^5_{3/2}$ & $3d_{5/2}$ & 1
& 9674.71(79)[09] & 38 & 9673.08 & 9673.11 & \\

& 3F
& $2s^2 2p^5_{1/2}$ & $3s_{1/2}$ & 1
& & & 10341.89 & & \\

& 3B
& $2s 2p^6_{1/2}$ & $3p_{1/2}$ & 1
& & & 10997.41 & & \\

& 3C
& $2s^2 2p^5_{1/2}$ & $3d_{3/2}$ & 1
& 11122.8(11)[04] & 14 & 11124.82 & 11123.68 & \\

& 3A
& $2s 2p^6_{1/2}$ & $3p_{3/2}$ & 1
& & & 11437.08 & & \\

& E2S
& $2s 2p^6_{1/2}$ & $3d_{5/2}$ & 2
& & & 11715.81 & & \\

\midrule

\textbf{Ir} ($Z=77$)
& M2
& $2s^2 2p^5_{3/2}$ & $3s_{1/2}$ & 2
& 9011.65(44)[15] & 32 & 9009.85 & 9009.13 & \\

& 3G
& $2s^2 2p^5_{3/2}$ & $3s_{1/2}$ & 1
& 9019.94(22)[15] & 63 & 9018.79 & 9018.08 & \\

& E2L
& $2s^2 2p^5_{3/2}$ & $3p_{1/2}$ & 2
& 9171.91(42)[16] & 20 & 9169.92 & & \\

& E2M
& $2s^2 2p^5_{3/2}$ & $3p_{3/2}$ & 2
& 9641.6(12)[07] & 5 & 9639.63 & & \\

& 3E
& $2s^2 2p^5_{3/2}$ & $3d_{3/2}$ & 1
& & & 9804.07 & & \\

& 3D
& $2s^2 2p^5_{3/2}$ & $3d_{5/2}$ & 1
& 9954.38(44)[04] & 82 & 9953.39 & 9953.33 & \\

& 3F
& $2s^2 2p^5_{1/2}$ & $3s_{1/2}$ & 1
& 10681.1(11)[22] & 11 & 10679.97 & 10679.58 & \\

& 3B
& $2s 2p^6_{1/2}$ & $3p_{1/2}$ & 1
& & & 11349.40 & & \\

& 3C
& $2s^2 2p^5_{1/2}$ & $3d_{3/2}$ & 1
&  &  & 11496.73 & 11495.57 & \\

& 3A
& $2s 2p^6_{1/2}$ & $3p_{3/2}$ & 1
& & & 11816.76 & & \\

& E2S
& $2s 2p^6_{1/2}$ & $3d_{5/2}$ & 2
& & & 12104.47 & & \\

\midrule

\textbf{Bi} ($Z=83$)
& M2
& $2s^2 2p^5_{3/2}$ & $3s_{1/2}$ & 2
& 10496.7(12)[04] & 42 & 10497.65 & 10496.37 & \\

& 3G
& $2s^2 2p^5_{3/2}$ & $3s_{1/2}$ & 1
& 10508.34(39)[45] & 133 & 10507.64 & 10506.38 & 10507.8(12) \\

& E2L
& $2s^2 2p^5_{3/2}$ & $3p_{1/2}$ & 2
& 10677.2(10)[05] & 26 & 10678.40 & 10677.07 & 10680.9(13) \\

& E2M
& $2s^2 2p^5_{3/2}$ & $3p_{3/2}$ & 2
& 11346.8(36)[11] & 8 & 11348.81 & 11347.62 & \\

& 3E
& $2s^2 2p^5_{3/2}$ & $3d_{3/2}$ & 1
& & & 11529.65 & & \\

& 3D
& $2s^2 2p^5_{3/2}$ & $3d_{5/2}$ & 1
& 11727.5(8)[15] & 34 & 11725.05 & 11724.22 & 11722.2(18) \\

& 3F
& $2s^2 2p^5_{1/2}$ & $3s_{1/2}$ & 1
& & & 12873.28 & & 12874.1(15) \\

& 3B
& $2s 2p^6_{1/2}$ & $3p_{1/2}$ & 1
& & & 13629.67 & & 13626.7(23) \\

& 3C
& $2s^2 2p^5_{1/2}$ & $3d_{3/2}$ & 1
& & & 13927.75 & & \\

& 3A
& $2s 2p^6_{1/2}$ & $3p_{3/2}$ & 1
& & & 14296.97 & & \\

& E2S
& $2s 2p^6_{1/2}$ & $3d_{5/2}$ & 2
& & & 14645.20 & & \\

\bottomrule
\end{tabular}
\end{table*}

\section{\label{theory}Theoretical calculations}

\subsection{Collisional-radiative model}
Collisional-radiative (CR) modeling is essential for interpreting spectra from plasmas created by a quasi-monoenergetic electron beam such as those produced in the NIST EBIT. The NOMAD code used in the present work is a non-Maxwellian, time-dependent CR model designed to simulate population kinetics and light emission from diverse plasmas \cite{ralchenko2001}. It solves coupled rate equations for level populations across multiple charge states, incorporating tens of thousands of atomic energy levels and millions of transition pathways. The input data for NOMAD (i.e., energies, transition probabilities, electron-impact cross sections) were calculated by the FAC relativistic atomic structure code~\cite{Gu2008}.

The NOMAD model accounts for electron-impact (de-)excitation and ionization, radiative decay, autoionization, and recombination processes. It is particularly suited for EBIT conditions, where the electron energy distribution is non-thermal. The code can simulate both steady-state and transient plasma conditions, making it valuable for analyzing time-resolved spectra and understanding population kinetics.

In our work, NOMAD was used to model the populations and spectra of Ne-like and nearby ions, including metastable states and forbidden transitions. The simulations provided insight into the relative level populations and transition intensities to aid the line identification as shown in Fig. \ref{fig:exp_theo}.

\subsection{MCDHF+RCI Calculations}

In this work, we employed the GRASP2K package, which implements the MCDHF method, to compute transition energies \cite{Jonsson2013, Jonsson2023}. The MCDHF technique is based on the variational principle, where atomic state functions are expressed as linear combinations of configuration state functions that share the same parity and total angular momentum. These CSFs are constructed from one-electron Dirac spinors, whose radial components and expansion coefficients are optimized through the self-consistent solution of the MCDHF equations.

Initial MCDHF calculations were performed for the reference configurations 1s$^2$2s$^2$2p$^6$, 1s$^2$2s$^2$2p$^5$3$l$ ($l$ = s, p, d), and 1s$^2$2s2p$^6$3$l$ for Ne-like ions. Electron correlation effects were subsequently included by expanding the CSF space using the active set approach. This involved single and double electron excitations from the $n$ = 2 and 3 shells of the reference configurations into active orbitals with maximum principal quantum numbers up to $n$ = 7 and orbital angular momentum quantum numbers $l = 0, 1, \dots, n-1$, with the active orbitals added layer by layer. Single excitations from the inner-shell 1s orbital were also included. These expansions account for both valence-valence and core-valence correlations and resulted in the convergence of the calculated energies to within 0.0001\% - 0.001\%. Following the self-consistent-field procedure, relativistic configuration interaction (RCI) calculations were performed using the same CSF expansions to include nuclear finite-mass corrections, the Breit interaction, and leading QED effects such as self-energy and vacuum polarization corrections.

\section{\label{analysis}Data Analysis}

Calibration lines were first identified manually in a reference pixel of the TES array and used as anchor points for aligning spectral features across all remaining pixels. The corresponding line positions in each pixel were determined using dynamic time warping \cite{szypryt_transitionedge_2019, salvador_accurate_2007}, which provides robust feature matching across the array despite small pixel-to-pixel variations in detector response. To correct for slow gain drift over the course of the measurement, a linear correction based on the quiescent current level of each pixel was applied following the entropy-minimization approach described in Ref.~\cite{fowler_practice_2016}. This correction was determined using pulses near the energy region of interest and resulted in a modest improvement in the overall full-width-at-half-maximum (FWHM) energy resolution.

\begin{figure}
\centering
\includegraphics[width=0.99\linewidth]{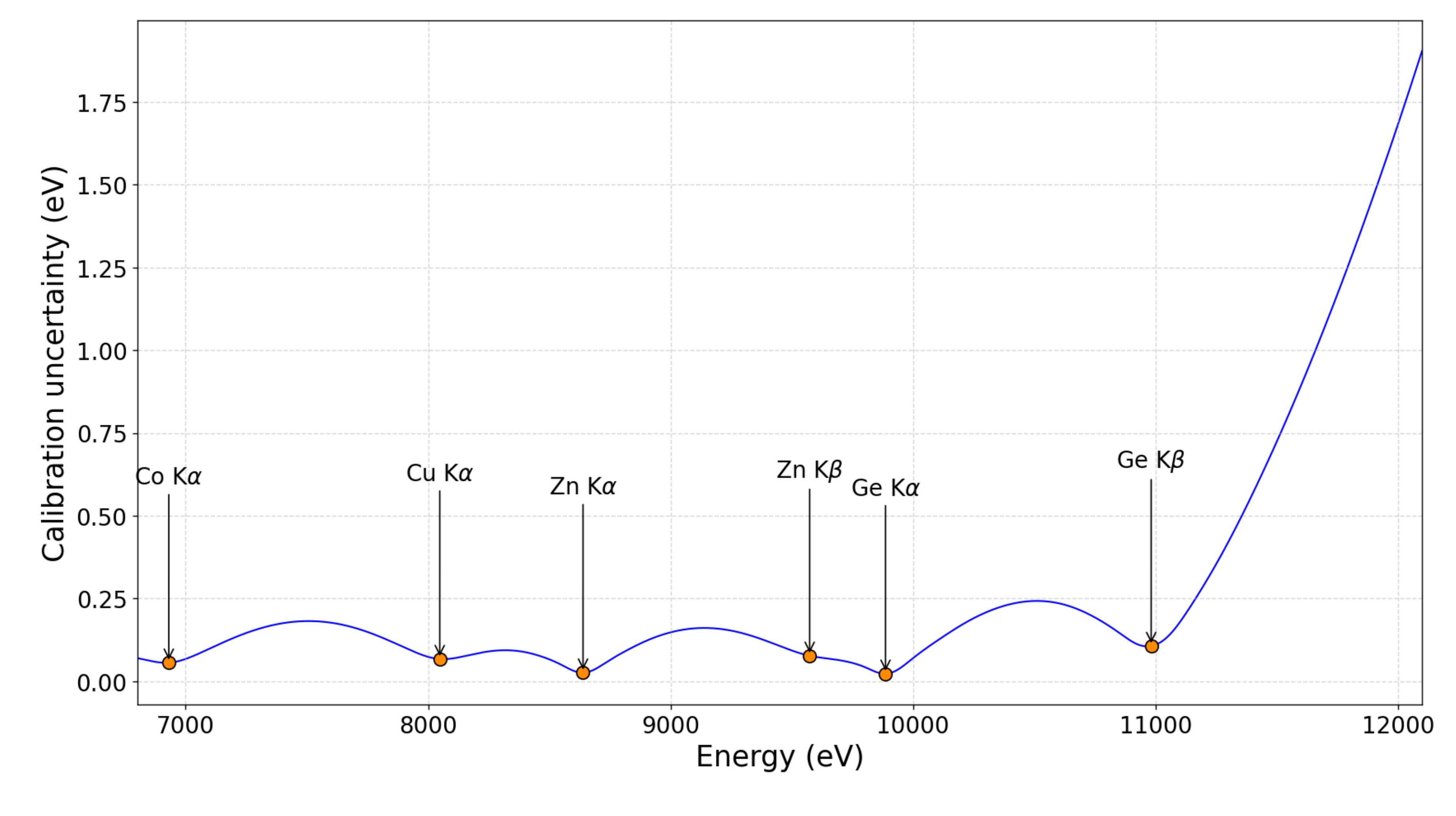}
\caption{Example of calibration uncertainties obtained through the Gaussian Process Regression analysis as a function of energy~\cite{fowler_energy_2022}. The positions of the fluorescence reference lines from the NIST SRD-128 \cite{Deslattes2003,SRD128} used in the TES calibration are indicated. The uncertainty is smallest near strong calibration features and increases in the interpolating regions between them.}
\label{calibrationuncert}
\end{figure}

Calibration of the TES array was performed using well-characterized fluorescence x-ray lines from metallic samples mounted on an external aluminum ring holder and excited by a commercial x-ray source in a dedicated vacuum chamber. The reference K$\alpha$ and K$\beta$ transitions from Co, Cu, Zn, and Ge span the energy range relevant for the present measurements as shown in Fig.~\ref{fig:calibration_lines}. The calibration source was turned on for one minute in every five, and has a significantly higher flux than the EBIT source. The locations of these calibration peaks were identified in each pixel and fit using previously measured line-shape models referenced to the SI energy scale. For each TES pixel, a calibration function was then used to relate pulse height to photon energy. Following Ref.~\cite{fowler_energy_2022}, the calibration function was constructed within the framework of Gaussian Process Regression (GPR). The GPR calibration model is mathematically equivalent to a spline constrained by a cost function that balances the $\chi^2$ goodness-of-fit against a curvature penalty, with the latter given by the integral of the square of the second derivative over the fit range \cite{wahba1978improper}. The regularization parameter governing this balance was determined by maximizing the marginal likelihood of the calibration data in a Bayesian framework \cite{fowler_energy_2022}. The result of the GPR is the calibration uncertainty and gives the square brackets in Table \ref{tab:ne_like_all}. For each thermal cycle of the cryostat a separate GPR derived calibration curve and associated uncertainty is produced, with one given as example in Fig. \ref{calibrationuncert}. Pixels that did not reproduce the expected calibration features or showed unstable behavior were excluded from further analysis. The final spectrum is co-added between the 72 working pixels that survived the cuts.

The GPR analysis yields a continuous calibration uncertainty function across the full energy range of the TES spectrum. The calibration uncertainty reaches a minimum at the positions of the strongest reference lines and increases smoothly between calibration points where interpolation dominates. The calibration uncertainty at the energy of each measured transition was obtained directly from this continuous uncertainty curve and propagated as a component of the total uncertainty budget.

After calibration, spectra from all valid TES pixels were co-added to maximize signal-to-noise. Spectral features were fit with Gaussian profiles to determine line centroids and their statistical uncertainties. To assess sensitivity to analysis choices, the fitting procedure was repeated for several spectral bin sizes. The standard deviation of the extracted line energies across these binning choices was taken as an estimate of systematic uncertainty associated with spectral binning and fitting procedure. Final uncertainties were obtained by summing the statistical, calibration, and systematic contributions in quadrature.

\begin{figure}[t]
\centering
 \includegraphics[width=0.99\linewidth]{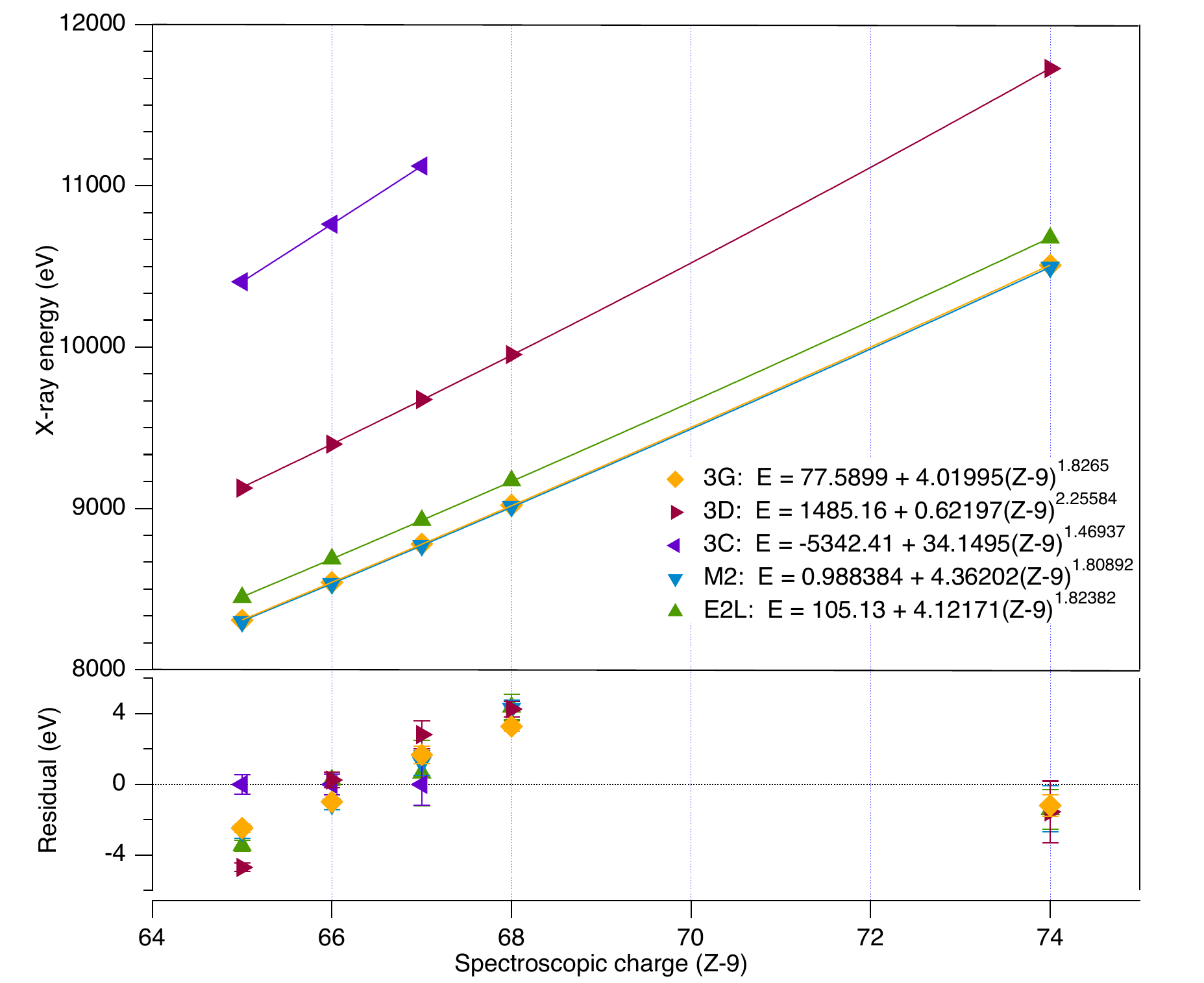}
\caption{x-ray transition energies along the Ne-like isoelectronic sequence versus spectroscopic charge, Z-9. Power-law fits providing approximate estimates of transition energies for the intervening elements. Fit residuals are shown in the lower panel.}
\label{fig:EnFit}
\end{figure}

\section{\label{results}Results}

High-resolution spectra of the measured transitions in Ne-like Re$^{65+}$, Os$^{66+}$, and Ir$^{67+}$ include magnetic quadrupole (M2), electric quadrupole (E2), and dipole-allowed (E1) lines, with sub-eV precision approximately 8~keV - 12~keV energy range presented in Table \ref{tab:ne_like_all}. Table~\ref{tab:ne_like_all} also includes transition energies measured in this work for Ne-like W$^{64+}$ and Bi$^{73+}$. The line identifications are supported by our CR model calculations using the NOMAD code that reproduces the relative intensity ratios of the measured spectra as illustrated in Fig. \ref{fig:exp_theo}. Uncertainties include statistical, fitting, and calibration components and are below 1 eV for the most prominent Ne-like transitions. The measured energies follow a smooth dependence on spectroscopic charge that can be represented by power-law fits, as shown in Fig.~\ref{fig:EnFit}, providing interpolated transition energies for the unmeasured intermediate elements. 

The TES detector enables the resolution of previously blended features, including satellite lines from adjacent charge states. Figure~\ref{fig:exp_theo} shows the recorded spectra, illustrating the separation of key Ne-like transitions in the measured elements. Satellite lines observed near strong E1 transitions are attributed to Na-like, Mg-like, and Al-like ions, as confirmed by NOMAD collisional-radiative modeling. These features, visible in the theoretical spectra (see Fig. \ref{fig:exp_theo}), arise from adjacent charge states and provide insight into the charge state distribution and excitation pathways within the EBIT plasma.

Table~\ref{tab:ne_like_all} lists our measured Ne-like W transition energies and their uncertainties alongside relativistic many-body perturbation theory (RMBPT) calculations~\cite{Safronova2001} and earlier crystal-spectrometer measurements~\cite{Beiersdorfer2012} (see also Fig.~\ref{fig:exp_theo}). Multireference M{\o}ller--Plesset (MRMP) calculations of Vilkas et al.~\cite{vilkas2008} provided the best agreement with the previously measured transition energies in W~\cite{Beiersdorfer2012}. For W, our M2, 3G, and 3D energies agree with the earlier measurements within the combined experimental uncertainties. Our E2L and 3F energies are closer to the MRMP values of Vilkas et al. but are about $3\sigma$ below the earlier measurements. The 3C energy is 3.01~eV ($4\sigma$) below the earlier measurement. The MRMP calculations do not cover the other elements measured in the present work.  

\section{\label{discussion}Discussion}

The TES measurements presented here extend the Ne-like sequence into a high-$Z$ regime beyond the region of avoided level crossings. In this domain, configuration mixing effects have largely stabilized, allowing the experimental data to provide benchmarks for future precision atomic structure calculations and tests of the QED contributions to the line energies. 

Comparison of the measured transition energies with the GRASP2K and RMBPT calculations reveals discrepancies of up to several eV. In particular, both theoretical approaches generally underestimate the energies of the prominent 3D transitions across the measured isoelectronsic sequence. The differences between GRASP2K range from approximately 1.0 eV for W and Ir to 2.5 eV for Bi, without an obvious monotonic dependence on atomic number. A similar trend is observed for the 3G transition, although agreement is in general better. Furthermore, neither theoretical approach consistently provides better agreement across all transitions. These differences demonstrate that reproducing transition energies to sub-eV accuracy remains challenging for current atomic structure calculations.

The GRASP2K energies for the 3D transition are consistently below the measured values, while the deviations for the M2 transition have both signs.



\section{\label{conclusions}Conclusions}

We have measured $n = 2$–$3$ transitions in Ne-like Re, Os, and Ir with sub-eV precision using TES microcalorimeter spectrometers. We also measured transition energies in Ne-like W and Bi, extending the earlier measurements for these ions. These results expand the experimental dataset for high-$Z$ Ne-like ions and offer new benchmarks for atomic structure calculations.

The spectra reveal well-resolved transitions and satellite features, demonstrating the effectiveness of the TES detector and calibration strategy. In particular, the use of a dedicated fluorescence-based calibration source enabled accurate energy measurements in the 8~keV to 12~keV region, which is critical for studies of K- and L-shell transitions in mid- to high-$Z$ elements. This calibration approach significantly extends the EBIT’s traceable measurement capabilities and ensures the reliability of the extracted line energies.

The measurements confirm trends observed in earlier studies and provide new data in a previously unexplored $Z$ range. Future work will focus on collisional-radiative models to better account for observed line intensities and population dynamics.

\begin{acknowledgments}
This work was funded by a NIST Grant Award Number 70NANB20H87 and by a National Science Foundation Award Number 2309273. Yu.R. was supported by NASA under award number 80GSFC21M0002.
\end{acknowledgments}

\bibliographystyle{apsrev4-2}
\bibliography{references}

\end{document}